\documentclass[fleqn,10pt]{wlscirep1}
\usepackage{lipsum}
\usepackage{graphicx}
\usepackage{algorithm}
\usepackage[noend]{algpseudocode}
\usepackage{enumerate}
\usepackage{url}
\usepackage{pgfplots}
\usepackage{soul}
\usepackage{subcaption}
\usepackage{xr}
\usepackage{epsfig}
\usepackage{array}
\usepackage[normalem]{ulem}

\usepackage{sidecap}

\title{Dynamic network analysis improves protein 3D structural classification}

\author[1]{Khalique Newaz}
\author[1]{Jacob Piland}
\author[2]{Patricia L. Clark}
\author[4]{Scott J. Emrich}
\author[3]{Jun Li}
\author[1,*]{Tijana Milenkovi\'{c}}
\affil[1]{Department of Computer Science and Engineering, University of Notre Dame, Notre Dame, IN 46556, USA}
\affil[2]{Department of Chemistry and Biochemistry, University of Notre Dame, Notre Dame, IN 46556, USA}
\affil[3]{Department of Applied and Computational Mathematics and Statistics, University of Notre Dame, Notre Dame, IN 46556, USA}
\affil[4]{Department of Electrical Engineering and Computer Science,  University of Tennessee, Knoxville, TN 37996, USA.}
\affil[*]{Corresponding author (email: tmilenko@nd.edu)}

\begin{abstract}
Protein structural classification (PSC) is a supervised problem of assigning proteins into pre-defined structural (e.g., CATH or SCOPe) classes based on the proteins' sequence or 3D structural features. We recently proposed PSC approaches that model protein 3D structures as protein structure networks (PSNs) and analyze PSN-based protein features, which performed better than or comparable to state-of-the-art sequence or other 3D structure-based approaches in the task of PSC. However, existing PSN-based PSC approaches model the whole 3D structure of a protein as a static PSN. Because folding of a protein is a dynamic process, where some parts of a protein fold before others, modeling the 3D structure of a protein as a dynamic PSN might further help improvethe existing PSC performance. Here, we propose for the first time a way to model 3D structures of proteins as dynamic PSNs, with the hypothesis that this will improve upon the current state-of-the-art PSC approaches that are based on static PSNs  (and thus upon the existing state-of-the-art sequence and other 3D structural approaches). 
Indeed, we confirm this on 71 datasets spanning ${\sim}$44,000 protein domains from CATH and SCOPe.
\end{abstract}
\begin{document}

\flushbottom
\maketitle

\thispagestyle{empty}

\section{Introduction}
\label{sec:introduction}
\subsection{Background and motivation}
Protein structural classification (PSC) uses sequence or 3D structural features of proteins and pre-defined structural classes to perform a supervised learning of a classification model, which can then be used to predict classes of currently unclassified proteins based on their features. Because  structural similarity of proteins often indicates their functional similarity, PSC can help understand functions of proteins. That is, the predicted structural class of a protein can be used to predict functions of the protein based on functions of other proteins that have the same class as the protein of interest. 

Traditional PSC approaches rely heavily on sequence-based protein features \cite{xia2016}. However, it has been argued that proteins with low (high) sequence similarity can have high (low) 3D structural similarity \cite{Sousounis2012, kosloff2008sequence}. Hence, 3D-structural features offer complementary insights to sequence features in the task of PSC.
Traditional 3D-structural protein features are extracted \emph{directly} from 3D structures of proteins \cite{cui2008classification}. In contrast, one can first model the 3D structure of a protein using a \emph{protein structure network (PSN)}, where nodes are amino acids and two nodes are joined by an edge when the corresponding amino acids are close enough to each other in the protein's 3D structure. Then, one can use PSN (i.e., network) features of proteins in the task of PSC. Modeling 3D structures of proteins as PSNs could help gain novel insights about protein folding because it opens up opportunities to apply an arsenal of approaches from the network science field to study 3D protein structures \cite{newazcodon}.

We already proposed a PSN-based PSC approach called NETPCLASS \cite{newaz2020} that performed better than or comparable to existing state-of-the-art sequence or other 3D structure-based  (i.e., non-PSN-based) approaches in the task of PSC, i.e., when classifying protein domains from two established databases, i.e., Class, Architecture, Topology, and Homology (CATH) \cite{greene2006cath} and Structural Classification of Proteins (SCOP) \cite{murzin1995scop}. Thus, PSN analysis is a state-of-the-art in the task of PSC.
NETPCLASS relies on ordered PSNs, where nodes in a PSN have order based on positions of the corresponding amino acids in the protein sequence.
Following NETPCLASS, we proposed an improved PSC approach that relies on weighted PSNs instead of ordered PSNs, where each edge in a PSN is assigned a weight that quantifies distances between amino acids in terms of both sequence and 3D structure \cite{guo2019}. Intuitively, a higher edge weight signifies that the two amino acids are close in the 3D space even though they are far apart in the sequence, while a low edge weight signifies that the two amino acids are close in the 3D space only because they are also close in the sequence.

Both of the above existing PSN-based PSC approaches rely on \emph{static} PSNs, which capture the final (i.e., native) 3D structure of a protein; henceforth, we refer to the above two PSN types as static unweighted ordered PSNs and static weighted unordered PSNs, respectively. However, folding of a protein is a temporal process, as some parts of the protein fold earlier than others before the protein acquires its native 3D structure \cite{Hu2013}.
Hence, using this \emph{dynamic} 3D structural information of a protein, which captures the 3D (sub)structural configurations of the protein as the protein undergoes folding to attain its native structure, can be more informative. The few experimental methods that can determine such information \cite{Cassaignau2016, waudby2013} are not scalable, i.e., are restricted to studying only a few 3D (sub)structural configurations of a single protein \cite{komar2018unraveling}.  So, experimental data on structures resulting from such a dynamic protein folding process are lacking. Because there is abundant availability of experimental data for the native structures of proteins, can we use this information to approximate the 3D (sub)structural configurations of proteins?
Existing computational, simulation-based methods do rely on the native 3D structure as a proxy to predict (sub)structures of a protein fold \cite{trovato,Zhao2020}. However, they are limited to a few proteins (e.g., nine \cite{Zhao2020}), and typically rely on small, model proteins, rather than a representative selection of proteins that appear in nature \cite{trovato}.

\subsection{Our contributions}
Here, we propose to approximate 3D (sub)structural configurations of proteins using \emph{dynamic} PSNs, with the hypothesis that this will improve PSC performance compared to static PSNs that only capture native 3D structures of proteins. 
If our proposed dynamic PSN-based PSC approach outperforms existing static PSN-based PSC approaches, then it would open up opportunities to explore protein folding-related research questions using dynamic, rather than static, PSNs. One such research question is to explore the effect of synonymous codon usage on co-translational protein folding \cite{walsh2020synonymous}, which has significance in heterologous protein expression for drug discovery \cite{gustafsson2004}. Because co-translational protein folding leads to partial folds of a protein during translation before the protein acquires its final 3D structure, a dynamic PSN-based analysis as opposed to the current static PSN-based analysis \cite{newazcodon} could provide more insights.

\begin{figure}[!h]
    \centering
    \includegraphics[scale=0.6]{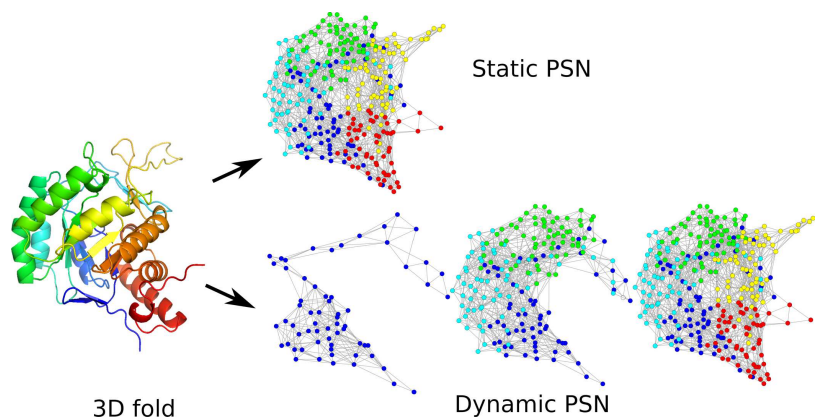}
    \caption{Illustration of the static vs. dynamic PSN of a protein 3D structure.}
    \label{fig:fig1}
\end{figure}

Given the 3D native structure of a protein, we model it as a dynamic PSN. Intuitively, our dynamic PSN consists of multiple PSN ``snapshots'', where the final snapshot captures the native 3D structure of the protein, while the earlier snapshots approximate 3D (sub)structural views of the protein as it attains its final 3D structure via the folding process (Fig. \ref{fig:fig1}). Specifically, given information about the native 3D structure of a protein, we convert it into a dynamic PSN by forming incremental cumulative groups of amino acids (i.e., sub-sequences) of the protein starting from the amino- or $N$-terminus (i.e., start of the protein sequence) up to the carboxyl- or $C$-terminus (i.e., end of the protein sequence). Then, we construct a PSN snapshot corresponding to each incremental sub-sequence. All snapshots collectively form a dynamic PSN of the protein. Currently, our dynamic PSNs are unweighted and unordered. 

Our proposed dynamic PSNs are dynamic versions of existing static unweighted unordered PSNs \cite{newaz2020}. Hence, to show that dynamic information helps in the task of PSC, it is sufficient to show that our dynamic PSNs outperform static unweighted unordered PSNs. Consequently, we compare our proposed dynamic PSNs with static unweighted unordered PSNs in the task of PSC. Additionally, we compare our dynamic PSNs with each of the other existing, more advanced notions of static weighted unordered PSNs \cite{guo2019} and static unweighted ordered PSNs \cite{newaz2020}, in the task of PSC. Note that we do not propose dynamic versions of the above two existing static PSN types (weighted and ordered) because of the lack of computational approaches that can extract meaningful information (i.e., features) from the corresponding dynamic versions. Instead, we propose a dynamic version of static unweighted unordered PSNs because there exist advanced computational approaches using which we can extract meaningful features from this dynamic version. We hope to explore in our future work both adding weights and node order to our dynamic PSNs as well as developing novel approaches for analyzing such dynamic weighted ordered PSNs.

To extract features from our proposed dynamic PSNs, we rely on  graphlets; small subgraphs of a large network \cite{newaz20195}. Specifically, we use the concepts of dynamic graphlets \cite{hulovatyy2015} and graphlet orbit transitions (GoTs) \cite{aparicio2018}. Despite having different mathematical definitions, intuitively, both dynamic graphlets and GoTs capture how local network neighborhoods of nodes of a dynamic network change over time.

To extract features from the considered existing PSN types, i.e., static unweighted unordered PSNs, static weighted unordered PSNs, and static unweighted ordered PSNs, we use the concepts of original graphlets \cite{prvzulj2004}, weighted graphlets \cite{guo2019}, and ordered graphlets \cite{malod2014gr}, respectively, which we again note have been shown to be among state-of-the-art features in the task of PSC \cite{newaz2020, guo2019}.

We compare our proposed dynamic PSNs (i.e., the corresponding features) with each of the existing static PSNs (i.e., the corresponding features) in the task of PSC. Note that we do not compare our proposed dynamic PSNs with other non-PSN-based protein features because we already showed that existing static PSNs perform better than or are comparable to existing state-of-the-art non-PSN-based protein features in the task of PSC \cite{newaz2020}. Hence, if we show that our proposed dynamic PSNs outperform existing static PSNs, then by transitivity, this would mean that our proposed dynamic PSNs also outperform existing non-PSN-based protein features.

Specifically, we evaluate the considered PSN types under the same classification algorithm, i.e., logistic regression (LR), for a fair comparison. For a given input protein, the output of an LR classifier is a set of likelihoods with which the protein belongs to the considered structural classes. Note that previously we explored other conventional classifiers, e.g., support vector machine, but this did not yield improvement, i.e., LR performed better in our considered task of PSC \cite{newaz2020}. Also, note that we use LR as opposed to a non-conventional (i.e., deep learning-based) classification algorithm because our proposed dynamic PSNs already perform very well under LR (see below), and because a non-conventional classification algorithm would also come at a cost of much larger computational time. However, recently, we observed an encouraging PSC performance boost when the static weighted unordered PSNs were used with a deeplearning-based algorithm compared to when they were used with a conventional algorithm \cite{guo2019}. So, we may test whether this holds for our proposed dynamic PSNs in our future work.

We evaluate the considered PSN types in the task of classifying ${\sim}$44,000 protein domains from CATH \cite{greene2006cath} and Structural Classification of Proteins--extended (SCOPe) \cite{scope} databases.  We transform protein domains to PSNs with labels corresponding to CATH and SCOPe structural classes, where we study each of the four hierarchy levels of CATH and SCOPe, resulting in 71 protein domain datasets. Our performance evaluation is based on quantifying the misclassification rate, i.e., the fraction of PSNs in the test data for which the trained models give incorrect structural class predictions, using 5-fold cross-validation.

Over all of the 71 considered datasets, our proposed dynamic PSNs statistically significantly outperform their static counterpart, i.e., static unweighted unordered PSNs, in the task of PSC. 
Additionally, our dynamic PSNs statistically significantly outperform the other PSN types, i.e., static weighted unordered and static unweighted ordered PSNs. 
Moreover, our dynamic PSNs not ``only'' outperform all of the existing PSN types, but also, for most of the datasets, their performance is exemplary. Namely, for 66 out of 71 (i.e., 93\%) of the datasets, our proposed dynamic PSNs show a misclassification rate of only $0.1$ or below.

\section{Methods}
\label{sec:methods}

\subsection{Data}
\label{data-description}
We use all of the 145,219 currently available Protein Data Bank (PDB) IDs \cite{berman2000} that have sufficient 3D crystal structure resolution (i.e., resolution values of 3 Angstrom (\AA) or lower). This set of PDB IDs contains 408,404 unique protein sequences (i.e., chains). To identify protein domains within these protein chains, we rely on two protein domain structural classification databases: CATH \cite{greene2006cath} and SCOPe \cite{scope}.

Because protein chains that are almost sequence identical can affect our downstream PSC analysis, as is typically done \cite{newaz2020}, we only keep a set of protein chains such that each chain in the set is less than 90\% sequence identical to any other chain in the set, using the procedure from Supplementary Section S1. This procedure results in 35,131 sequence non-redundant protein chains, where each chain has at least one CATH- or SCOPe-based protein domain, resulting in a total of 60,434 CATH and 25,864 SCOPe protein domains.

Both CATH and SCOPe classify protein domains based on four hierarchical levels of protein 3D structural classes. The lower the hierarchy level of a class, the higher the structural similarity of protein domains within the class. Because we aim to perform supervised protein structural classification at each hierarchy level, including the fourth, to have enough statistical power \cite{figueroa2012predicting}, we only consider those fourth-level CATH and SCOPe protein structural classes where each class has at least 30 protein domains. Thus, we only consider those protein domains that belong to any one such class, which results in 34,791 CATH and 9,394 SCOPe domains.

For each of the 34,791 CATH and 9,394 SCOPe domains, we create a static unweighted unordered protein structure network (PSN) (Section \ref{psn-construction}). Then, as is typically done \cite{newaz2020}, we only keep those domains whose corresponding PSNs have a single connected component. Additionally, we keep only those domains that have at least 30 amino acids. This results in 34,630 CATH and 9,329 SCOPe domains. These are the final sets of CATH and SCOPe domains that we use in our study.

Given all of the 34,630 CATH domains, we test the ability of the considered features to distinguish between the first-level classes of CATH, i.e., \emph{mainly alpha} ($\alpha$), \emph{mainly beta} ($\beta$), and \emph{alpha/beta} ($\alpha$/$\beta$), which have at least 30 domains each. Hence, we consider all 34,630 CATH domains as a single dataset, where the protein domains have labels corresponding to three first-level CATH classes: $\alpha$, $\beta$, and $\alpha/\beta$. Second, we compare the features on their ability to distinguish between the second-level classes of CATH, i.e., within each of the first-level classes, we classify domains into their sub-classes. To ensure enough training data, we focus only on those first-level classes that have at least two sub-classes with at least 30 domains each. All three first-level classes satisfy this criteria. For each such class, we take all of the domains belonging to that class and form a dataset, which results in three datasets. Third, we compare the approaches on their ability to distinguish between the third-level classes of CATH, i.e., within each of the second-level classes, we classify domains into their sub-classes. Again, we focus only on those second-level classes that have at least two sub-classes with at least 30 domains each. 16 classes satisfy this criteria. For each such class, we take all of the domains belonging to that class and form a dataset, which results in 16 datasets. Fourth, we compare the approaches on their ability to distinguish between the fourth-level classes of CATH, i.e., within each of the third-level classes, we classify PSNs into their sub-classes. We again focus only on those third-level classes that have at least two sub-classes with at least 30 domains each. 28 classes satisfy this criteria. For each such class, we take all of the domains belonging to that class and form a dataset, which results in 28 datasets.

Thus, in total, we analyze 1+3+16+28=48 CATH datasets. We follow the same procedure for SCOPe and obtain 1+6+5+10=22 SCOPe datasets. In addition to the above 48+22=70 CATH and SCOPe datasets, as is typically done \cite{newaz2020}, we use an additional dataset  because of the following reason. Typically, high sequence similarity of proteins indicates their high 3D structural similarity. Consequently, given a set of proteins, if proteins that belong to the same 3D structural class have high sequence identity (typically $>40\%$), then a sequence-based protein feature might be sufficient to compare them \cite{Burkhard1999}. So, we aim to evaluate how well our considered PSN (i.e., 3D structural) features can identify protein domains that belong to the same structural class when all of the domains (within and across structural classes) come from protein sequences that show low ($\leq 40\%$) sequence identity. To do this, we use the Astral dataset from the SCOPe database that has 14,666 protein domains, where the label indicates the protein family (i.e., the fourth-level structural class) to which the domain belongs. We follow the same filtering criteria for the Astral dataset that we do above for the CATH and SCOPe domains, which results in a single dataset with 729 domains and 18 structural classes. Hence, including all of the CATH, SCOPe, and Astral datasets, in total, we use 71 datasets in this study.
 
\subsection{PSN types and their corresponding features}
\label{psn-construction}
For each protein domain, we construct four types of PSNs, out of which three are existing notions of static PSN types and one is our proposed notion of dynamic PSN type. The three existing static PSN types cover all of the types of PSNs that currently exist in the literature. These existing PSN types are \emph{(i)} static unweighted unordered PSN (named \emph{original}), \emph{(ii)} static weighted unordered PSN (named \emph{weighted}), and \emph{(iii)} static unweighted ordered PSN (named \emph{ordered}). The fourth PSN type, i.e., our proposed dynamic PSN (named \emph{dynamic}),  is the dynamic version of the static unweighted unordered PSN type (Table \ref{tab:psn_type}). 

\begin{table}[!ht]
\centering
    \caption{Summary of the four considered PSN types, i.e., our proposed dynamic PSN type and three existing static PSN types. Each row (except the first) corresponds to a PSN type, and each column (except the first) outlines the corresponding PSN characteristic. Note that original PSNs differ from dynamic PSNs in a single aspect and the two are thus fairly comparable. On the other hand, weighted PSNs and ordered PSNs differ from dynamic in two aspects. }
    \label{tab:psn_type}
    \begin{tabular}{|c|c|c|c|}
    \hline
    & Static & Unweighted &  Unordered  \\
    PSN type & or & or &  or \\
    & dynamic & weighted &  ordered \\
    \hline
    \textbf{Dynamic}&  Dynamic &  Unweighted  &  Unordered\\ 
      \hline
      \textbf{Original}& Static  &  Unweighted & Unordered \\
      \hline
      \textbf{Weighted}&  Static &  Weighted &  Unordered  \\ 
      \hline
     \textbf{Ordered}& Static & Unweighted &  Ordered\\ \hline
\end{tabular}
\end{table}

Below, we provide details of how we create each PSN type, as well as how we extract graphlet features from each PSN type. The latter is explained only briefly in the main paper due to space constraints, and it is explained in detail in Supplementary Section S2.
\vspace{0.1cm}
\newline
\noindent \textbf{Existing static unweighted unordered PSNs (i.e., original).} Given a protein domain, we first obtain the corresponding Crystallographic Information File (CIF) file from the PDB \cite{berman2000} that contains information about the 3D coordinates of the heavy atoms (i.e., \emph{carbon}, \emph{nitrogen}, \emph{oxygen}, and \emph{sulphur}) of the amino acids in the domain. Then, we create the static unweighted unordered PSN for the given domain using established criteria \cite{faisal2017, newaz2020, guo2019,milenkovic2009}: \emph{(i)} we consider amino acids as nodes, and \emph{(ii)} we define an edge between two amino acids if the spatial distance between any of their heavy atoms is within 6{\AA}. To extract features from this PSN type, we rely on edge-based graphlet degree vector matrix (eGDVM) \cite{solava2012}, which is based on the concept of original graphlets \cite{prvzulj2004}. For details, see Supplementary Sections S2 and S3.
\vspace{0.1cm}
\newline
\noindent \textbf{Existing static weighted unordered PSNs (i.e., weighted).} Given a protein domain, we first create the static unweighted unordered PSN as described above. Then, using an established approach \cite{guo2019},  we assign weights to edges of the PSN, where an edge weight is the geometric mean of two types of distances between the corresponding amino acids, i.e., their sequential distance and the inverse of their spatial distance. To extract features from the resulting static weighted unordered PSN type, we rely on an existing measure called weighted edge-based graphlet degree vector matrix (weGDVM), which is based on the concept of weighted graphlets and has been successfully used in the task of PSN-based PSC \cite{guo2019} (Supplementary Section S2).
\vspace{0.1cm}
\newline
\noindent \textbf{Existing static unweighted ordered PSNs (i.e., ordered).} Given a protein domain, we first create the static unweighted unordered PSN as described above. Then, using an established approach \cite{malod2014gr}, we add a node order to the PSN, such that the node order represents the positions of the nodes (i.e., amino acids) in the sequence of the corresponding protein domain. To extract features from the resulting static unweighted ordered PSN type, we rely on an existing measure called ordered graphlet feature vector (oGFV), which is based on the concept of ordered graphlets \cite{malod2014gr} and has been successfully used in the task of PSN-based PSC \cite{newaz2020} (Supplementary Section S2).
\vspace{0.1cm}
\newline
\noindent \textbf{Our proposed dynamic unweighted unordered PSNs (i.e., dynamic).} Given a protein domain, starting from the first amino acid (i.e., the amino acid closest to the $N$-terminus) of the domain sequence, we define multiple incremental sub-sequences of the protein domain, where each sub-sequence $i+1$ includes all of the amino acids from the previous sub-sequence $i$. Then, for each sub-sequence, i.e., using the amino acids in the given sub-sequence, we create a static unweighted unordered PSN using the same procedure as we do in subsection ``Existing static unweighted unordered PSN (i.e., original)''. We identify this PSN corresponding to a protein sub-sequence as a PSN snapshot. The collection of all PSN snapshots corresponding to all of the sub-sequences for the given domain forms the dynamic unweighted unordered PSN of that domain. We create two types of dynamic unweighted unordered PSNs using two distinct strategies to create PSN snapshots, i.e., to define protein domain sub-sequences, as follows.
\vspace{0.1cm}
\newline
\noindent\underline{Strategy 1:} Given a protein domain of length $S$ (i.e., containing $S$ amino acids), we create $S/5$ sub-sequences (i.e., PSN snapshots) as follows. Starting from the first amino acid of the sequence, the first snapshot has first 5 amino acids, the second snapshot has first 10 amino acids, the third snapshot has first 15 amino acids, and so on, until we include all of the $S$ amino acids of the protein domain in the last PSN snapshot. We use an increment of five amino acids per snapshot because it intuitively mimics addition of individual 3D secondary structural elements (i.e., $\alpha$-turns or $\beta$-strands that are usually 3 to 7 amino acids long) as the domain folds into its final 3D structure. 
\vspace{0.1cm}
\newline
\noindent\underline{Strategy 2:} With the above strategy, different domains might have different numbers of PSN snapshots. However, it might be beneficial to consider the same number of PSN snapshots over all domains, for consistency. So, here, given all of the protein domains that we analyze, we first identify the length of the smallest domain. We find that the smallest domain is 30 amino acids long. Second, we define PSN snapshots of the smallest domain using the same approach as described above in strategy 1, which results in six PSN snapshots. Then, for each of the other considered protein domains, we create a dynamic PSN with six PSN snapshots. That is, given a protein domain, we define incremental snapshots in the increments of $100/6$ ${\sim}$ $17\%$ of amino acids in the corresponding protein domain. 
\vspace{0.1cm}
\newline
To extract features of dynamic PSNs created using each of the above two strategies, we rely on two existing measures, i.e., dynamic graphlet degree vector matrix (dGDVM) \cite{hulovatyy2015} and graphlet orbit transition matrix (GoTM) \cite{aparicio2018}. Thus, we consider $2\times2=4$ combinations of dynamic PSNs and their features. For more details, see Supplementary Section S2.

\subsection{The classification framework}
\label{lr_framework}
For each of the 71 datasets (Section \ref{data-description}), we train a logistic regression (LR) classifier for each of the considered PSN features (Section \ref{psn-construction}). 
Specifically, for a given dataset and a given feature, we perform 5-fold cross-validation. That is, first, we divide the dataset into five equal-sized folds (or subsets), where in each subset we keep the same proportion of different protein structural classes as present in the initial dataset. Second, for each such subset, we use the PSNs in the subset as the test data and the union of PSNs in the remaining four subsets as the training data. 

But before we train an LR model using a training data and use it to predict classes of PSNs in a test data, we use the training data itself in a 5-fold cross validation manner \cite{kohavi1995study, salzberg1997comparing} to perform hyper-parameter tuning, i.e., to choose  an ``optimal'' value for the regularization hyper-parameter. We perform linear search on 10 equally-spaced log-scaled values between $2^{-8}$ and $2^8$ to find an optimal value. After we choose the ``best'' hyper-parameter value, we use it and all of the training data to train an LR model (Supplementary Section S4).

After training an LR model, we evaluate its performance on the test data using misclassification rate -- the percentage of all PSNs from the test data that are not classified into their correct protein structural classes. We compute both aggregate and average misclassifiation rate. Aggregate misclassification rate is a single misclassification rate over all five folds, while average misclassification rate is the per-fold misclassification rate averaged over the five folds.  For any of the considered features, for any of the considered datasets, we find negligible differences between the two measures. Hence, for each feature, for each dataset, we only report the corresponding aggregate misclassification rate for simplicity.

\section{Results and discussion}

We hypothesize that modeling protein 3D structures using dynamic, as opposed to static, PSNs will capture more information about the 3D structural organisation of proteins in the task of PSC. 
To confirm this hypothesis, it is necessary and sufficient for our proposed dynamic unweighted unordered PSNs to outperform existing static unweighted unordered PSNs. 
If additionally our dynamic PSNs outperform the other, more advanced notions of static weighted unordered PSNs and static unweighted ordered PSNs (Section \ref{psn-construction}), then that would only further strengthen the importance of dynamic PSN-based modeling of protein 3D structures.
For each PSN type, we extract graphlet-based features from the corresponding PSNs (Section \ref{psn-construction}) and use the features in the task of PSC (Section \ref{lr_framework}). 
Because we consider four combinations of dynamic PSN construction strategies and features (Section \ref{psn-construction}), in Section \ref{sec:best_proposed}, we first evaluate the four combinations to choose the best one. Then, we compare the best combination, i.e., the best dynamic PSN approach, to original PSNs (Section \ref{sec:best_proposed_vs_best_existing}), weighted PSNs (Section \ref{sec:best_proposed_vs_others}), and ordered PSNs (Section \ref{sec:best_proposed_vs_others}), in the task of PSC. Finally, in Section \ref{sec:dynamicpsn_summary}, we summarize the overall performance of dynamic PSNs compared to all existing PSN types combined.

\subsection{Selection of the best dynamic PSN approach}
\label{sec:best_proposed}
Recall that we use four combinations of dynamic PSN construction strategies and features (or four dynamic PSN approaches): strategy 1 with dGDVM, strategy 2 with dGDVM, strategy 1 with GoTM, and strategy 2 with GoTM (Section \ref{psn-construction}). We evaluate each dynamic PSN approach on each of the 71 considered datasets in the task of PSC and obtain the corresponding $71$ misclassification rates (Section \ref{lr_framework}). Then, we compare all four dynamic PSN approaches to each other to check whether a given dynamic PSN approach outperforms each of the other three dynamic PSN approaches. 
Specifically, for each pair of approaches, we evaluate whether the misclassification rates of the given approach are significantly better (i.e., have lower values) than the corresponding misclassification rates of the other approach, using paired Wilcoxon rank sum test. 
Because we perform three pairwise comparisons for each of the four approaches, in total, we perform $3\times4=12$ pairwise comparisons. 
Thus, we obtain 12 $p$-values, which we correct using False Discovery Rate (FDR) (\cite{benjamini1995}) to obtain the corresponding adjusted $p$-values (i.e., $q$-values). We find that the dynamic PSN approach based on strategy 1 with dGDVM significantly ($q$-value $< 10^{-8}$) outperforms each of the other three dynamic PSN approaches (Supplementary Fig. S1). Henceforth, we report results only for this dynamic PSN approach.

\subsection{Dynamic PSNs outperform original PSNs}
\label{sec:best_proposed_vs_best_existing}

\begin{figure}[!h]
    \centering
    \includegraphics[scale=0.9]{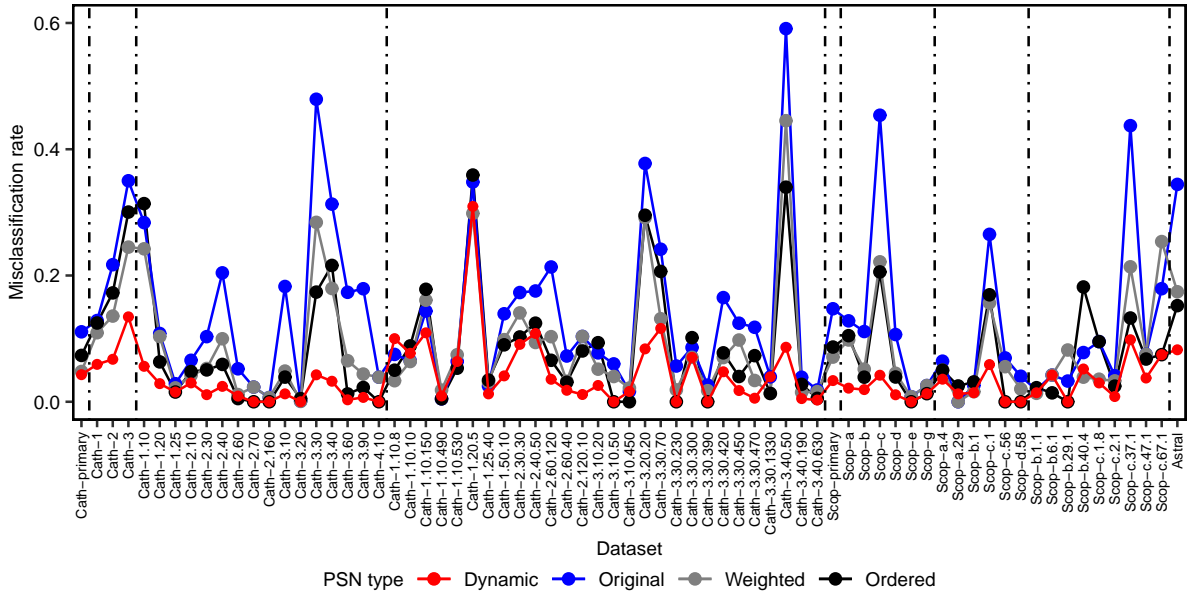}
    \caption{Misclassification rates of all four PSN types for each of the $71$ datasets (48 from CATH, 22 from SCOPe, and Astral; Section \ref{data-description}). In red are results for our proposed dynamic PSNs. In blue, gray, and black are results for the existing static PSN types.}
    \label{fig:final}
\end{figure}

\begin{figure}[!h]
    \centering
    \includegraphics[scale=0.9]{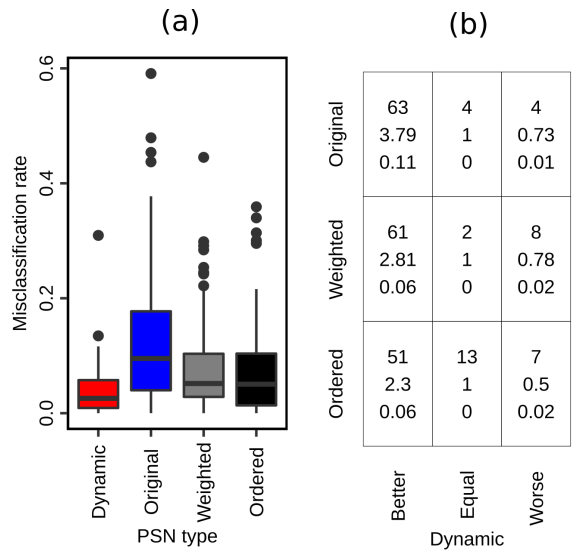}
    \caption{Performance summary of the four PSN types over all 71 datasets. Panel (a) compares distributions of the 71 misclassification rates between the four PSN types. Panel (b) compares dynamic PSNs ($x$-axis) with each of the original, weighted, and ordered PSNs ($y$-axis). In a given cell in panel (b), the three numbers mean the following. The top number indicates the number of datasets in which dynamic PSNs perform better than, the same as, or worse than the corresponding original, weighted, or ordered PSNs. The middle number indicates the relative increase (if greater than 1) or decrease (if less than 1) in the performance of dynamic PSNs compared to original, weighted, or ordered PSNs. The bottom number indicates the absolute increase or decrease in the performance of dynamic PSNs compared to original, weighted, or ordered PSNs. }
    \label{fig:combo}
\end{figure}

To check whether dynamic PSNs outperform original PSNs or vice versa, we compare the two PSN types by comparing their misclassification rates over all 71 datasets. That is, we check whether misclassification rates of dynamic PSNs are better than original PSNs and vice versa, using the paired Wilcoxon rank-sum test (Section \ref{sec:best_proposed}). Hence, we obtain two $p$-values, which we correct using FDR to obtain the corresponding $q$-values. We find that dynamic PSNs significantly ($q$-value of $5.22\times 10^{-12}$) outperform original PSNs (Figs. \ref{fig:final} and \ref{fig:combo} (a)).

In more detail, dynamic PSNs are superior to original PSNs for 63 of the 71 (i.e., ${\sim}88\%$) datasets. The two PSN types are tied for four of the 71 (i.e., ${\sim}6\%$) datasets. Dynamic PSNs are inferior to original PSNs for only four of the 71 (i.e., ${\sim}6\%$) datasets (Fig. \ref{fig:combo} (b)). Thus, the 71 datasets can be partitioned into three groups, depending on whether dynamic PSNs perform better than, the same as, or worse than original PSNs.
Given a group of datasets, we quantify the increase or decrease in performance of dynamic PSNs compared to original PSNs using \emph{relative change} and \emph{absolute change} between misclassification rates of the two PSN types. To compute relative change, we divide original PSNs' average misclassification rate over all considered datasets with dynamic PSNs' average misclassification rate over all considered datasets. Hence, a relative change of more than one means that dynamic PSNs are superior to original PSNs. A relative change of less than one means that dynamic PSNs are inferior to original PSNs. A relative change of one means that the two PSN types perform the same. To compute absolute change, we take the absolute difference between original PSNs' average misclassification rate over all considered datasets and dynamic PSNs' average misclassification rate over all considered datasets. 
We find that dynamic PSNs are superior to original PSNs with a relative increase of $3.79$ and an absolute increase of $0.11$ in performance, while dynamic PSNs are inferior to original PSNs with a relative decrease of $0.73$ and an absolute decrease of $0.01$ in performance (Fig. \ref{fig:combo}(b)). That is, dynamic PSNs are not only superior for many more datasets than original PSNs, but also, dynamic PSNs improve more on average upon original PSNs than the latter improve on average upon the former.

In particular, it is important to note that for the Astral dataset, dynamic PSNs show a relative increase of $4.18$ in performance over original PSNs, decreasing the misclassification rate from $0.344$ to $0.082$, which is close to ``ideal'' performance (i.e., to misclassification rate of 0). This is important because all domains in the Astral dataset come from a set of proteins with low ($\leq40\%$) sequence identities (Section \ref{data-description}). Hence, the likelihood of a sequence-based protein feature to work well for such dataset is low, and proposing a good 3D structure- or PSN-based protein feature for such dataset is important.

\subsection{Dynamic PSNs outperform weighted PSNs and ordered PSNs}
\label{sec:best_proposed_vs_others}

Next, we compare our dynamic PSNs, which are unweighted and unordered, against each of the remaining two existing static PSN types, i.e., weighted PSNs and ordered PSNs. We perform each comparison (dynamic PSNs against weighted PSNs and dynamic PSNs against ordered PSNs) the same way we have performed the comparison of our dynamic PSNs against original PSNs in Section \ref{sec:best_proposed_vs_best_existing}. We find that dynamic PSNs significantly outperform both weighted PSNs ($q$-value of $6.44\times 10^{-10}$) and ordered PSNs ($q$-value of $9.20\times10^{-9})$.

Similar to the way we quantify the performance increase or decrease of dynamic PSNs against original PSNs in Section \ref{sec:best_proposed_vs_best_existing}, we use relative and absolute changes to
quantify the performance increase or decrease of dynamic PSNs against weighted PSNs and against ordered PSNs. Regarding performance comparison of dynamic PSNs against weighted PSNs, dynamic PSNs are superior to weighted PSNs for 61 of the 71 (i.e., ${\sim}86\%$) datasets, with a relative increase of $2.81$ and an absolute increase of $0.06$ in performance. The two PSN types are tied for two of the 71 (i.e., ${\sim}3\%$) datasets. Dynamic PSNs are inferior to weighted PSNs for only eight out of the 71 (i.e., ${\sim}11\%$) datasets, with a relative decrease of $0.78$ and an absolute decrease of $0.02$ in performance (Fig. \ref{fig:combo} (b)). Regarding performance comparison of dynamic PSNs against ordered PSNs, dynamic PSNs are superior to ordered PSNs for 51 of the 71 (i.e., ${\sim}72\%$) datasets, with a relative increase of $2.3$ and an absolute increase of $0.06$ in performance. The two PSN types are tied for 13 of the 71 (i.e., ${\sim}18\%$) datasets. Dynamic PSNs are inferior to ordered PSNs for only seven out of the 71 (i.e., ${\sim}10\%$) datasets, with a relative decrease of $0.5$ and an absolute decrease of $0.02$ in performance (Fig. \ref{fig:combo} (b)). That is, dynamic PSNs are not only superior for many more datasets than weighted and ordered PSNs, but also, dynamic PSNs improve more on average upon weighted and ordered PSNs than the latter two improve on average upon the former.

Additionally, similar to Section \ref{sec:best_proposed_vs_best_existing}, dynamic PSNs show a decent performance boost for the Astral dataset compared to each of the weighted PSNs and ordered PSNs. That is, for the Astral dataset, dynamic PSNs show a relative increase of $2.12$ in performance over weighted PSNs and a relative increase of $1.85$ in performance over ordered PSNs, decreasing the misclassification rate of $0.174$ for weighted PSNs and of $0.152$ for ordered PSNs to $0.082$.

\subsection{Summary of the performance of dynamic PSNs}
\label{sec:dynamicpsn_summary}
So far, we compared dynamic PSNs against \emph{each} existing PSN type \emph{individually}.
Here, we compare dynamic PSNs against \emph{all} of the existing PSN types \emph{at the same time}, in order to evaluate the performance of dynamic PSNs against the best of the existing PSN types for each of the 71 datasets. Consequently, 
we find that dynamic PSNs perform strictly better than all existing PSN types for 44 datasets, while they are tied with the best of the existing PSN types for 14 datasets. That is, for $44+14=58$ of the 71 (i.e., ${\sim}82\%$) datasets, dynamic PSNs perform the best. For the remaining $13$ of the 71 (i.e., only ${\sim}18\%$) datasets, at least one of the existing static PSN types is better than dynamic PSNs.

Given the above two groups of datasets (the 58 datasets where dynamic PSNs are the best and the 13 datasets where dynamic PSNs are not the best), we examine whether any of the two groups are biased towards any of the four hierarchy levels of CATH or SCOPe. We do this to understand whether dynamic PSNs are more likely to perform better or worse for datasets from a certain hierarchy level of CATH or SCOPe. Specifically, given a group of datasets and a hierarchy level of CATH or SCOPe, we evaluate whether the given group contains a statistically significantly high number of (i.e., is enriched in) datasets from the given hierarchy level, and we do so by using the hypergeometric test \cite{falcon2008} (Supplementary Section S5). Because there are two groups of datasets and $4+4=8$ CATH and SCOPe hierarchy levels, in total, we perform 16 hypergeometric tests and obtain the corresponding 16 $p$-values. We correct the $p$-values using FDR to obtain the corresponding $q$-values. We find that the group of datasets where dynamic PSNs perform the best is statistically significantly ($q$-values $< 0.05$) enriched in each of the four hierarchy levels of CATH and SCOPe, or equivalently that the group of datasets where dynamic PSNs do not perform the best is not enriched in any of the hierarchy levels. This shows that dynamic PSNs perform consistently well for all four hierarchy levels of CATH and SCOPe. 

\vspace{-0.3cm}
\section{Conclusion}
We propose to capture the 3D structural organisation of a protein using a dynamic PSN, which approximates the 3D (sub)structural configurations of a protein as the protein attains its native 3D structure. 
We hypothesize that our dynamic PSNs should capture more of 3D structural information than existing static PSNs in the task of protein structural classification (PSC), where the latter have already been established as a state-of-the-art compared to traditional sequence and 3D structural PSC approaches.
We evaluate our hypothesis on protein domains and corresponding 3D structural labels from CATH and SCOPe databases. We show that our dynamic PSNs significantly outperform all existing static PSN types in the task of PSC. Our results are thus expected to guide future development of PSC approaches. Additionally, because we show that dynamic PSNs capture more complete 3D structural information of proteins than static PSNs, our study opens up opportunities to explore protein folding-related research questions using dynamic, rather than static, PSNs.

\section*{Funding}
This work is funded by a grant from the National Institutes of Health (R01 GM120733).

\bibliography{ref}




\end{document}